\documentstyle[emulapj]{article}

\font\tenbg=cmmib10 at 10pt

\def \rvecmu{{\hbox{\tenbg\char'026}}}

\font\tenbg=cmmib10 at 10pt

\def \rvecmu{{\hbox{\tenbg\char'026}}}

\begin{document}



\title{MHD Simulations of Bondi Accretion 
to a Star in the
 ``Propeller" Regime}

\author{M.M.~Romanova}
\affil{Department of Astronomy,
Cornell University, Ithaca, NY 14853-6801;
romanova@astrosun.tn.cornell.edu}

\author{O.D. Toropina}
\affil{Space Research Institute, Russian Academy
of Sciences, Moscow, Russia;
toropina@mx.iki.rssi.ru}

\author{Yu.M. Toropin}
\affil{
CQG International Ltd.,
\\ 10/5 Sadovaya-Karetnaya,
Build. 1 103006, Moscow, Russia; yuriy@cqg.com 
}

\author{R.V.E.~Lovelace}
\affil{Department of Astronomy, Cornell University, Ithaca,
NY
14853-6801; RVL1@cornell.edu }

\begin{abstract}

     This work investigates   Bondi
accretion  to a rotating magnetized star in
the ``propeller" regime using axisymmetric
resistive, magnetohydrodynamic simulations.
    In this regime accreting matter tends
to be expelled from the equatorial region
of the magnetosphere where the centrifugal
force on matter rotating with the star
exceeds the gravitational force.
   The regime is predicted to occur
if the magnetospheric radius larger
than the corotation radius and less
than the light cylinder radius.
    The simulations show that accreting
matters is expelled from the equatorial
region of the magnetosphere and that it
 moves away from the star in a 
supersonic, disk-shaped outflow.
    At larger radial distances the outflow
slows down and becomes subsonic. 
   The equatorial matter outflow is initially
driven by the centrifugal force, but
at larger distances the pressure gradient becomes 
significant. 
   We find that the star is spun-down mainly by
the  magnetic torques  at its surface with
the rate of loss of angular momentum $\dot{L}$
proportional to $-\Omega_*^{1.3}\mu^{0.8}$, where
$\Omega_*$ is the star's rotation rate
and $\mu$ is its magnetic moment.
   Further, we find that $\dot{L}$ is approximately
independent of the magnetic diffusivity
of the plasma $\eta_m$.
    The fraction of
the Bondi accretion rate which
accretes to the surface
of the star is found to be
$\propto \Omega_*^{-1.0}\mu^{-1.7}\eta_m^{0.4}$.
   Predictions of this work are important
for the observability of isolated old neutron
stars and for wind fed pulsars in
X-ray binaries.

\end{abstract}

\keywords{accretion, dipole
--- plasmas --- magnetic
fields --- stars: magnetic fields ---
X-rays: stars}

\section{Introduction}

  Rotating magnetized neutron
stars pass through different stages in
their evolution (e.g.,
 Shapiro \& Teukolsky 1983;  Lipunov 1992).
  Initially, a rapidly rotating
($P \lesssim 1{\rm s}$) magnetized
neutron star is expected to be
active as a radiopulsar.
   The star spins down owing to
the wind of magnetic field and
relativistic particles from the
region of the light cylinder $r_L$
(Goldreich and Julian 1969).
   However, after the
neutron star spins-down sufficiently,
the light cylinder radius
 becomes larger than magnetospheric
radius $r_m$ where the ram
pressure of external matter equals
the magnetic pressure in the
neutron star's dipole field.
  The relativistic wind
is then suppressed by the
inflowing matter (Shvartsman 1970).
    The external matter may come from
the wind from a binary companion or
from the interstellar medium
for an isolated neutron star.
    The centrifugal force
in the equatorial region at $r_m$ is much larger than
gravitational force if $r_m$ is much
larger than the corotation radius
$r_{cor}$.
    In this case the  incoming
matter tends to be flung away from
the neutron star by its
rotating magnetic field.
  This is the so called
``propeller" stage of evolution
(Davidson \& Ostriker 1973;
Illarionov \& Sunyaev 1975).

The ``propeller" stage of evolution,
though important, is  still
not well-understood theoretically.
   Different  studies have found
different dependences for the
spin-down rate of the star
(Illarionov \& Sunyaev 1975;
Davis, Fabian, \& Pringle 1979; Davis
\& Pringle 1981; Wang \& Robertson 1985; 
Lipunov 1992).
  Observational signs of the propeller
stage have been discussed by  
number of authors (e.g., Stella,
White, \& Rosner 1986; Treves \& Colpi 1991; 
Cui 1997; Treves et
al. 2000).

MHD simulations of  disk accretion 
to a rotating star in the
propeller regime  were done by 
Wang and Robertson (1985).
However, authors considered only
equatorial plane and concentrated
on investigation of instabilities
at the boundary between the
magnetosphere and surrounding medium.
Thus they could not
investigate accretion along the
magnetic poles of the star.
     An analytical model of disk
accretion in the propeller regime
was developed by Lovelace,
Romanova, and Bisnovatyi-Kogan (1999).
 Disk accretion at the stage
of weak propeller ($r_m \sim r_{cor}$)
investigated numerically by Romanova
et al. (2002).


\begin{figure*}[t]
\centering
\caption{Geometry of the MHD simulation
region, 
where $\dot{M}_B$ is  the Bondi accretion
rate, $\rvecmu$ and ${\bf \Omega}_*$ are
the magnetic moment and angular velocity
of the star, $R_*$ is the radius of
the star, $(R_{max}, Z_{max})$ are the limits
of the computational region.   
In the described calculations
$R_* \ll R_{max}$.
} 
\label{Figure 1}
\end{figure*}

  The mentioned  studies
obtained  possible trends of the
propeller stage of evolution.
  However, 
two and three dimensional
MHD simulations are needed to
obtain more definite answers to the
important physical questions.
    The questions which need to be
answered are:
   (1) What are the
physical conditions of the matter
flow around the star in the propeller
regime of accretion?
  (2) What is the spin-down rate
of the star, and how does it
depend on the star's
magnetic moment and rotation rate
and on the inflow rate
of the external matter?
  (3) What is the accretion rate  to the
surface of the star and how does it depend
on the star's rotation rate and magnetic moment?
   (4) What are the possible observational
consequences of this stage of evolution?

   This paper  discusses results of
axisymmetric, two-dimensional,
resistive MHD simulations of accretion
a rotating magnetized star 
in the ``propeller"
regime.
   We treat the case when matter
accretes spherically with the Bondi accretion rate.
 Bondi accretion to a non-rotating and a slowly
rotating star was investigated by 
Toropin et al. (1999; hereafter T99). 
    It was shown that the magnetized star 
accretes matter at a rate less than the
Bondi rate.
   Toropina et al. (2002; hereafter T02) 
confirmed this result for stronger magnetic
fields. 
     In this paper we
consider initial conditions similar to those in T02, but
investigate the case of rapidly rotating stars.
   Section 2 gives a rough physical treatment
of the propeller regime.
   Section 3 describes the numerical model
and computations.
  Section 4 describes the main results from
our simulations, and Section 5 gives a numerical
application of our results.
    Section 6 gives the conclusions from this work.

\section{Physics of the ``Propeller" Regime}

    In the propeller regime the magnetospheric
radius $r_m$ is larger than the corotation
radius $r_{cor}=(GM/\Omega_*^2)^{1/3}\approx
0.78\times 10^9 P_{10}^{2/3}$ cm and 
incoming matter in the equatorial plane is
flung away from the star.
   Here, we assume $M=1.4M_\odot$, and
$P_{10}=P/10$ s with $P$ the rotation
period of the star.
     The magnetospheric radius is
determined by a balance of the ram pressure of
the inflowing matter $p_m+\rho_m {\bf v}_m^2$
against the magnetic pressure ${\bf B}^2/8\pi$ 
of the neutron
star's field $B_m=\mu/r_m^3$, where the
$m-$subscripts indicate that the quantity is
evaluated at a distance $r_m$, and $\mu$ is
the star's magnetic moment.

   Consider first the case of a non-rotating
star.   
     The Bondi inflow velocity
at $r_m$ is supersonic for 
specific heat ratios $\gamma < 5/3$.
   Thus $\rho_m = \dot{M}/(4 \pi r_m^{3/2} \sqrt{GM})$,
where $\dot{M}$ is the  accretion rate. 
   For this case the ram pressure is 
approximately $\rho_m GM/r_m$ so that
$$
r_{m0} = \left({\mu^2 \over 
2 \dot{M} \sqrt{GM}} \right)^{2/7}~,
$$
$$
\approx 6.1\times 10^{10} 
\left({\mu_{30}^2 \over
\dot{M}_{-17}}\right)^{2/7}~{\rm cm} 
\eqno(1)
$$
(Davidson \& Ostriker 1973; Shapiro \& Teukolsky 1983).
 Here,
$\mu_{30}=\mu/10^{30}~{\rm G cm}^3$, 
and $\dot{M}_{-17}=\dot{M}/10^{-17}M_\odot/$yr.

    For a rapidly rotating star, $r_m \gg r_{cor}$, 
the dominant velocity is
due to the rotation of the magnetosphere so that
${\bf v}_m^2 = (\Omega_* r_m)^2$. The flow
velocity is much larger than the sound speed
so that this is a ``supersonic propeller''
(Davies et al. 1979).
   Neglecting the complicated two-dimensional
nature of the plasma flow, we  estimate
$\rho_m(\Omega_* r_m)^2 \approx \mu^2/(8\pi
r_m^6)$, which gives
$$ 
r_{m\Omega} \approx \left({\mu^2
\sqrt{GM} \over 2 \dot{M}
\Omega_*^2} \right)^{2/13} 
$$ 
$$
 \approx 0.81\times 10^{10}\left({\mu_{30}^2
P_{10}^2 \over \dot{M}_{-17}}
\right)^{2/13}~{\rm cm}~ 
\eqno(2) 
$$
(Wang \& Robertson 1985;  Lovelace, Romanova,
\& Bisnovatyi-Kogan 2002). 
    In this limit, there is to
a first approximation no accretion to the
star;  all of the incoming matter is flung
away from the star in the
equatorial plane by the  rotating magnetic
field.  

Equations (1) and (2) give
the same values for
$\Omega_{2} = (2\dot{M}/\mu^2)^{3/7}
(GM)^{5/7}$ which corresponds to a period
$P_{2}\approx 7100
(\mu_{30}^2/\dot{M}_{-17})^{3/7}$ s.
The propeller regime requires periods
$P < P_{2}$.
    At the other extreme, the condition
that $r_{m\Omega}$ be less than the
light cylinder radius gives the
condition $P> P_1\approx
0.77(\mu_{30}^2/\dot{M}_{-17})^{2/9}~{\rm s}$.

   The spin-down rate of the star can be
estimated as $Id\Omega_*/dt \approx 4\pi r_m^2
(r_mB_m^2/8\pi)$, which is the magnetic
torque at the radius $r_m$.  This assumes
that $B_\phi \approx B_m$ at $r_m$.
  Thus
$$ 
I { d\Omega_* \over dt}
\approx - {\mu^2 \over 2 r_{m\Omega}^3}=
-{\mu^{14/13} \Omega_*^{12/13} \dot{M}^{6/13}
\over 2^{7/13}(GM)^{3/13}}~.
\eqno(3) 
$$
This gives a spin-down time scale 
$$
\tau = {\Omega_* \over |\dot{\Omega}_*|}
\approx {2.2\times 10^7 ~{\rm yr}  \over
\mu_{30}^{14/13} \dot{M}_{-17}^{6/13}P_{10}^{1/13}}  ~,
$$
where  the neutron star's moment of
inertia $I$ is assumed to be $10^{45}$ g~$\!$cm$^2$.  
The angular momentum lost by the star
goes into the outflow of material mainly in the
equatorial plane.  
The rotational power lost by
 the star $I\Omega_* d\Omega_*/dt$ goes 
into the kinetic (thermal plus
flow) energy of the equatorial outflow.

\section{Numerical Model}

   We simulate the plasma flow in
the propeller regime using
 an axisymmetric, resistive 
MHD code.
    The code incorporates the methods 
of local iterations and 
flux-corrected-transport (Zhukov, 
Zabrodin, \& Feodoritova 1993).
  The code is described in our
earlier  investigation of 
Bondi accretion to a non-rotating
magnetized star (T99). 
    The equations for resistive MHD are
$$
  {\partial \rho \over
  \partial t}+
  {\bf \nabla}{\bf \cdot}
\left(\rho~{\bf v}\right)
=0{~ ,}
\eqno(4) 
$$ 
$$ 
\rho\frac{\partial{\bf v}} {\partial t}
            +\rho ({\bf v}
\cdot{\bf \nabla}){\bf v}=
  -{\bf \nabla}p+
{1 \over c}{\bf J}\times {\bf B} +
{\bf F}^{g}{ ~,} 
\eqno(5) 
$$ 
$$
  \frac{\partial {\bf B}}
{\partial t}
=
  {\bf \nabla}{\bf \times}
\left({\bf v}{\bf \times} {\bf B}\right)
  +
  \frac{c^2}{4\pi\sigma}
\nabla^2{\bf B} {~,} 
\eqno(6) 
$$ 
$$
  \frac{\partial (\rho\varepsilon)
}{\partial t}+
  {\bf \nabla}\cdot \left(\rho
\varepsilon{\bf v}\right)
=
  -p\nabla{\bf \cdot}
{\bf v} +\frac{{\bf J}^2} {\sigma}{~.} 
\eqno(7) 
$$
We assume axisymmetry $(\partial/\partial \phi =0)$,
but calculate all three components of ${\bf v}$
and ${\bf B}$.
  The equation of state is taken
to be that for an ideal gas,
$p=(\gamma-1)\rho
\varepsilon$, with specific heat ratio
$\gamma=7/5$.
    A value of $\gamma$ less than $5/3$
is expected in the case of an ionized
gas because of the influence of the
electron heat conduction.
    The equations incorporate Ohm's law ${\bf
J}=\sigma({\bf E}+{\bf v}
 \times {\bf B}/c)$, where $\sigma$ is
the electrical conductivity. 
   The associated magnetic diffusivity,
$\eta_m \equiv c^2\!/(4\pi\sigma)$, 
is considered to be 
a constant within the computational region.
    In equation (5) the gravitational force,
${\bf F}^{g} = -GM\!\rho{\bf
R}/\!R^3$, is due to the central star.

   The star rotates with angular 
velocity ${\bf \Omega}_*=\Omega_* ~ \hat{\bf z}$.
It is useful to introduce the dimensionless quantity
$\omega_* \equiv \Omega_*/\Omega_{K*} \leq 1$, where
$\Omega_{K*} =(GM/R_*^3)^{1/2}$ is the Keplerian angular
velocity at the stellar radius $R_*$.
   Simulations have been done for different
values of $\omega_*=0-0.7$.
   The intrinsic magnetic field of the 
star is taken to be an aligned dipole,
${\bf B} =[3{\bf R}\left({\rvecmu}\cdot
{\bf R} \right)-R^2 {\bf
\rvecmu} ]/{R^5}$ with 
$\rvecmu =\mu ~\hat{\bf z}$ and vector-potential
${\bf A}=\rvecmu \times{\bf R}/{R^3}$.

  We use a cylindrical, inertial coordinate
system $\left(r,\phi,z\right)$ 
with the $z-$ axis parallel to the
star's dipole moment $\rvecmu$ and  
rotation axis ${\bf \Omega}$.
    The equatorial plane
is treated as symmetry plane. 
   The vector potential $\bf A$ is
calculated so that ${\bf \nabla}\cdot{\bf
B}=0$ at all times.  The full set of
equations is given in T99.

\begin{figure*}[t]
\centering
\caption{
Matter flow in the ``propeller" regime
for a star rotating at $\Omega_*=0.5\Omega_{K*}$ after
$6.9$ rotation periods
of the star. 
   The axes are measured in units of the
star's radius. 
The background represents the
density and the length of
the   arrows is proportional to 
the poloidal velocity.
The thin solid lines
are magnetic field lines.
} 
\label{Figure 2}
\end{figure*}

To convert our equations to 
dimensionless form we measure
length in units of the Bondi radius
$R_B \equiv {GM}/c_\infty^2$, with
$c_\infty$ the sound speed
at infinity, density
in units of the density at infinity
$\rho_\infty$, and magnetic field strength
in units of $B_0$ which is the field
at the pole of the numerical star
($r=0, z=Z_*$). 
   Pressure is measured in units of 
$B_0^2/8\pi$.  
    The magnetic moment is measured in
units of $\mu_0 =B_0 R_B^3/2$.
   We measure velocity in units of
the Alfv\'en speed $v_0=B_0/\sqrt{4\pi \rho_\infty}$.
However, in the plots we give the velocities in units
of  $c_\infty$.

After putting equations (4)-(7)
in  dimensionless
form, one finds three
dimensionless parameters,
two of which are
$$
\beta \equiv \frac{8\pi
p_\infty}{B_0^2}, \quad \quad
\tilde{\eta}_m \equiv {\eta_m \over
R_B v_0} = {1\over S}{~.} 
\eqno(8)
$$ 
Here, $\tilde{\eta}_m$ is the
dimensionless magnetic diffusivity
which may be interpreted as
the reciprocal of the Lundquist 
number $S$; and
$p_\infty = \rho c_\infty^2/\gamma$ 
is the pressure at
infinity.
  The third parameter is $g={G
M_*}/({L_0 v_0^2}) = {\gamma
\beta  / 2}$ is of the order of $\beta$.
    We also use a parameter $b_0$
to control the field strength of
the star.  It is defined by the equation ${\bf
A}=b_0{\bf A_0}$, where 
$b_{0}$ has been varied in
the range $1 - 25$.

   Simulations were done in a cylindrical
``can''
$(0\le z\le Z_{max},~ 0\le r\le R_{max})$.
   The size of the  region was taken to be
$R_{max}=Z_{max}=R_s/\sqrt{2}=
R_{B}/5\sqrt{2}\approx 0.141R_B$, which is
less then the sonic radius of the Bondi
flow $R_{s}=[(5-3\gamma)/4]R_{B}=0.2R_B$ ($\gamma=7/5$).
   Thus matter inflows supersonically to
the computational region. 
    The inflow rate is taken to be the
 Bondi (1952) accretion rate,
$\dot{M}_B=4\pi\lambda(GM_{\star})^2
\rho_{\infty}/c_\infty^3$, where $\lambda=0.625$
for $\gamma=7/5$.  
    The incoming matter is assumed to be
unmagnetized.
    A uniform $(r,z)$ grid
with $N_R\times N_Z$ cells was used.
  For the results discussed here
$N_R \times N_Z=513 \times 513$.

    The  ``numerical star'',
is taken to be a cylinder
of radius $R_*$ and half axial length $Z_*$
with $R_*=Z_* <<R_{max},~ Z_{max}$. 
   In the results presented here,
$R_*=Z_*=0.0044 R_B$.  
   Clearly, the numerical star is very large:
the value of $R_*/R_B$,
although much less than unity,  is
much larger than the astrophysical ratio
for a neutron star ($\sim 10^{-4}$).
   The grid size, 
$DR=R_{max}/N_R=2.76\times10^{-4}R_B$,
is about $16$ times smaller than the star's radius.
   The vector-potential
${\bf A}$ of the intrinsic dipole field 
of the star is  determined 
on the surface of the numerical
star.
    The value of the vector potential
on this surface is fixed corresponding
to the numerical star being a perfect
conductor.   Equivalently, the component of the ${\bf B}$
field normal to the surface is fixed,
but the other two components vary.

      The density and
internal energy at the inner boundary 
(the numerical star) are set equal to
small numerical values.
   Thus, incoming matter
with higher values of these parameters,
is absorbed by the star. 
   This treatment of the star is
analogous to that used by Ruffert (1994).
    Typically, the
density of incoming matter is 
$\gtrsim 10^3$ times larger than  density
inside the star $\rho=1$.
    For a
purely hydrodynamic spherical 
accretion flow, we verified that
this accretor absorbs all incoming
matter at the Bondi rate.

\section{Matter Flow in the Propeller
Regime}

   Here, we first discuss 
simulations for the case $\omega_*=0.5$
$\beta=10^{-7}$, magnetic diffusivity
$\tilde{\eta}_m=10^{-5}$ (equation 8),
and for field strength $b_0=10$.
   At the end of this section we discuss
the dependence of the results on $\omega_*$,
$\mu$ (and $b_0$), and
${\eta}_m$. 

   Time is measured in units of the free-fall
time from the top or the side of the simulation
region, $t_{ff}=Z_{max}/v_{ff}$, with 
$v_{ff}=\sqrt{2GM/Z_{max}}$.  
   For the rotating star, time is also
given in terms of rotation periods of
the star, $P_*=2\pi/\Omega_*$.

    For the mentioned values of
$\beta$, $b_0$, and $\eta_m$, 
the magnetospheric
radius $r_{m0}$ and corotation radius
$r_{cor}=(GM/\Omega_*^2)^{1/3}$ are
equal for $\omega_* \approx 0.16$. 
    For smaller angular velocities, $\omega_* <
0.16$, the matter flow around the star is
close  to that in the non-rotating case.
   For $\omega_* > 0.16$ the flow exhibits
the features expected in
the ``propeller" regime.

 Figures 2 and 3 show the general nature  of
the flow in the propeller regime. 
   Two distinct regions separated by
a shock wave are observed:  
   One is the {\it external}
region where  matter inflows with the
Bondi rate and the density and
velocity agree well with the
Bondi (1952) solution.
   The second is the
{\it internal} region, where the flow is
strongly influenced by the stellar
magnetic field and rotation.
The shock wave, which divides these
regions, propagates outward as
in the non-rotating case (T99).
  For a rotating star in the propeller
regime the shock wave has the
shape of an ellipsoid flattened along
the rotation axis of the star.

   The region of the flow well within
the shock wave is approximately
time-independent.  The accretion rate
to the star becomes constant after
about $1-2$ rotation periods of the
star.

\begin{figure*}[t]
\centering
\caption{
Enlarged view of  Figure 2.
   The bold line represents the Alfv\'en surface.
Dotted line shows sonic surface.
The axes are measured in units  of  stellar
radii $R_*$.
} 
\label{Figure 3}
\end{figure*}

\begin{figure*}[t]
\centering
\caption{
Same as on Figure 3,
but the background represents angular velocity
$\Omega=v_\phi(r,z)/r$.
The axes are measured in units  of stellar
radii $R_*$.
}
 \label{Figure 4}
\end{figure*}

\begin{figure*}[b]
\centering
\caption{The left-hand panel (a) shows the
radial variation at $z=0$ of the 
azimuthal velocity $v_\phi$, the
radial velocity $v_r$, 
the sound speed $c_s$, and the escape
velocity $v_{esc}$, all in units
of $c_\infty$. 
  The right-hand panel (b) shows
the variation of the velocities along the
$z-$axis.  The fact that the sign of
the flow velocity changes across the
shock is due to shock's outward velocity.}
\label{Figure 5}
\end{figure*}

\begin{figure*}[t]
\centering
\caption{
Same as on Figure 3, but
bold solid lines are added which show 
the streamlines of the
matter flow.} 
\label{Figure 6}
\end{figure*}

\begin{figure*}[b]
\caption{
The figure shows the radial dependence
of the radial forces
acting to matter in the 
equatorial plane.
   The vertical scale is arbitrary.
The centrifugal force is the dominant 
one which drives matter to
the centrifugal equatorial wind.} 
\label{Figure 7}
\end{figure*}

\begin{figure*}[t]
\caption{
Angular velocity of
matter $\Omega=v_\phi/r$ versus $r$ 
in the equatorial region
for different angular velocities of the star.
    The dashed line represents the Keplerian
angular velocity $\Omega_K =\sqrt{GM/r^3}$. } 
\label{Figure 8}
\end{figure*}


\begin{figure*}[b]
\centering
\caption{ The left-hand panel ($a$) shows the
fraction of Bondi accretion rate reaching
the star as a function of the star's
angular velocity $\omega_*=\Omega_*/\Omega_{K*}$.
   For a non-rotating star,
$\dot{M}/\dot{M}_B =0.07$ or $\log(\dot{M}/\dot{M}_B)
=-1.1$.
   The right-hand panel ($b$) shows the 
dependence on the magnetic
moment $\mu$.   The straight lines show
the approximate
power law dependences.  
} 
\label{Figure 9}
\end{figure*}


\begin{figure*}[t]
\centering
\caption{
Total angular momentum outflow from the star
as a function of $\omega_*=\Omega_*/\Omega_{K*}$. 
   The straight line is a power law
fit $\dot{L} \propto - (\Omega_*)^{1.3}$.
This is somewhat steeper than the predicted
dependence of equation (3), 
$\dot{L} \propto - (\Omega_*)^{12/13}$, which
is shown by the dashed line in the figure.
   The vertical scale is arbitrary.
} 
\label{Figure 9}
\end{figure*}

   A new regime of matter flow forms
inside the expanding shock wave. 
  The rapidly
rotating magnetosphere expels
matter outward in the equatorial
region. 
   This matter flows
radially outward forming a low-density
rotating torus. 
   The outflowing matter
is decelerated when it reaches  
the shock wave.
   There, the flow changes direction 
and moves towards  the rotation axis of
the star.
 However, only a small fraction of this
matter accretes to the surface of
the star. 
  Most of the matter is
expelled again in the equatorial
direction by the rotating field
of the star.
    Thus, most of the 
matter circulates inside this
inner region driven by the rapidly
rotating magnetosphere. 
   Large-scale vortices 
form above and below the
equatorial torus (see Figures 2
and 3). 
In three dimensions
vortices of smaller scale may
also form. Thus, at the propeller
stage the significant part of the
rotational energy of the star may
go into the directed and thermal
energy of the expelled plasma.

    The solid line in Figure 3 shows the
Alfv\'en surface, where the magnetic
energy-density equals the thermal
plus kinetic energy-density,
${\bf B}^2/{8\pi}=\varepsilon + \rho
{\bf v}^2/2$. 
   In the equatorial plane, the Alfv\'en radius is
$r_A \approx 4.5R_*$. 
   For $r < r_A$, the magnetic pressure
dominates, and the magnetic field
is approximately that of a dipole.
   At larger distances the field 
is stretched by the outflowing
plasma. 
    Matter inside the
magnetopause (the region of closed magnetic
field lines inside the Alfv\'en
surface) rotates with the angular
velocity of the star (see Figure 4). 
    For $r > r_A$, matter continues
to rotate in the equatorial
region but the angular velocity
decreases with $r$. 
    Figure 5a shows the radial variation
of the velocities in the equatorial plane.
   Matter is accelerated by the rotating
magnetosphere for
$r > 4 R_*$ up to $r\sim 10
R_*$, but  at larger distances
the radial velocity decreases. 
   The azimuthal velocity
$v_\phi$ is significantly
larger than the radial velocity
at $r \sim 4R_*$, but they become
equal at larger distances. 
  The sound speed $c_s$ is  high
inside the shock wave
owing to the  Joule
heating ${\bf J}^2/\sigma$.
  Nevertheless, the
outflow is supersonic in the
region $r\sim (4 - 7) R_*$. 
  Note, that above and below 
the equatorial plane the
region of the supersonic outflow
is larger (see sonic surface line in
the Figures 3 and 4). 
    Figure 5b shows the variation of the 
velocities along the $z-$axis.

    Figure 6 shows
the streamlines of  the matter flow. 
   Matter free falls  along the
field lines going into the poles of
the star.
   Some matter which flows
close to the  $z-$axis 
accretes to the surface of the
star. 
   However,  matter more
removed  from the $z-$axis comes
close to the star, it is
deflected by the rotating magnetic
field, and it then moves outward in
the equatorial plane.

   Figure 7 shows the radial
dependences of the
different radial forces in
the equatorial plane. 
   One can see that for
$r>3.5 R_*$,  the
centrifugal force becomes dominant
in accelerating the 
matter outward. 
    However, at larger
distances, $r > 5.5 R_*$, the pressure
gradient force become larger and
determines acceleration of matter.
    Thus, centrifugal and pressure
gradient forces accelerate matter
in the radial direction. Note,
that in most of the region ($r >
4.3 R_*$) the magnetic force is
negative so that it opposes the matter
outflow.

    The results we have shown are for
a relatively strong propeller. 
   If the rotation rate is smaller,
then the matter outflow become less
intense. 
   For $\omega_* < 0.16$, no
equatorial outflow is observed
and the shock wave becomes
more nearly spherical as it
is in the non-rotating case.
    On the other hand the
shock wave becomes more a
more flattened ellipsoid for
larger $\omega_*$. 
  Matter rotates rigidly inside the
Alfv\'en radius $r_A$ while
at large distances  $\Omega= v_\phi/r$
decreases.
   Figure 8 shows the radial dependences of
the angular velocity in the
equatorial plane for different values
of $\omega_*$.

   Figure 9 shows that only 
a small  fraction
of the Bondi accretion rate 
accretes to the surface of the star. 
   Figure 9a shows that
the accretion rate to
the star decreases as the
angular velocity of the star increases,
$\dot{M}/\dot{M}_B \approx 0.0075~\!
\omega_*^{-1.0}$. 
   For a non-rotating star,
$\dot{M}/\dot{M}_B \approx 0.07$ in
agreement with our earlier work (T99).
   Figure 9b shows that the accretion
rate to the star decreases as 
the star's magnetic moment increases,
$\dot{M}/\dot{M}_B \sim \mu^{-1.7}$.
    We have also done a 
number of simulation runs for different
magnetic diffusivities in the range 
$\tilde{\eta}_m = 10^{-6}-10^{-4.5}$,
and from this we conclude
that $\dot{M}/\dot{M}_B \propto
({\eta}_m)^{0.4}$.
   Accretion along the rotation axis
of stars in the propeller regime was
discussed by Nelson, Salpeter, and Wasserman
(1993).
   Three-dimensional instabilities
(e.g., Arons \& Lea 1976) not included
in the present two-dimensional simulations
may act to increase the accretion rate
to the the star's surfaces.
 
   Figure 10 shows the total
angular momentum loss rate from
the star as a function of $\omega_*$.
This was obtained by evaluating the
integral
$$
\dot{L} = -\int d{\bf S}\cdot 
\left(\rho{\bf v}_pr v_\phi
-{ {\bf B}_p r B_\phi \over 4\pi }\right)~,
\eqno(9)
$$
over the surface of the numerical star.  
Here, $d{\bf S}$ is the outward pointing surface
area element and the $p-$ subscript indicates the
poloidal component.
   The dominant contribution to equation (9)
is the magnetic field term, and
this term increases relative to
the matter term as $\Omega_*$
increases.
   The dependence predicted by equation (3),
$\dot{L} \propto - \Omega_*^{12/13}$
is shown by the dashed line.
   The fact that the observed dependence,
$\dot{L} \propto - \Omega_*^{1.3}$ is steeper
may be due to the fact that the 
magnetospheric radius is not much larger
than the corotation radius.
   Also, we find that $\dot{L} \propto - \mu^{0.8}$
for a factor of four range of $\mu$ about our
main case.
   From simulation runs with
magnetic diffusivities in the range 
$\tilde{\eta}_m = 10^{-6}-10^{-4.5}$,
we conclude that $\dot{L}$ is approximately
independent of ${\eta}_m$.

\section {Astrophysical Example}

    Here, we give the 
conversions of the dimensionless
variables to physical values
for  the  case 
$\beta=10^{-7}$, $b_0=10$, and
$\tilde\eta_m=10^{-5}$ (equation 8).  
  We consider
accretion to a  magnetized
star with  mass
$M=1.4~M_{\odot}=2.8
\times 10^{33}$ g. 
   The density of
the ambient interstellar matter
is taken to be
 $\rho_\infty=1.7 \times
10^{-24}$ ~g$/$cm$^3$
($n_\infty=1/$~cm$^{3}$).
   We consider a higher than
typical sound speed in the ISM,
$c_\infty=100$ km/s, in order
to have a smaller Bondi radius
compatible with our simulations.
  The Bondi radius is
$$
R_B =GM_*/c_\infty^2 \approx 
1.9\times 10^{12}c_{100}^{-2}
~{\rm cm}~,
\eqno(10)
$$
 and the  Bondi (1952) 
accretion rate is
$$
\dot{M}_B=4 \pi \lambda
(GM_{NS})^2
 \rho_\infty/c_\infty^3
  \approx 0.7 \times
10^{-17}~(n_1/c_{100}^3){\rm
M}_{\odot}/{\rm yr}~,
\eqno(11)
$$
where $\lambda=0.625$ for $\gamma=1.4$,
$n_1 = n_\infty/1/$cm$^3$, and
$c_{100}=c_\infty/(100~ {\rm km/s})$.
    Using the definitions
$\beta=8\pi p_\infty/B_0^2$,
$p_\infty=\rho_\infty c_\infty^2/\gamma$,
and the fact that $b_0=10$,
we obtain
 the magnetic field at the
surface of the ``numerical"  star
$B_* \approx 1.7$ G.

   Recall that the size of the
numerical star is
 $R_*= 0.0044 R_B\approx 0.82
\times 10^{10}{\rm cm}$
in all simulations.
    Thus the magnetic moment
of the numerical star is
$\mu =B_* r_*^3/2\approx 0.48 
\times 10^{30}$ Gcm$^3$.
   Note that this is of the
order of the magnetic moment
of a neutron star with
surface field $10^{12}$ G and
radius $10^6$ cm.

    The numerical star rotates
with angular velocity
$\Omega_*=\omega_* 
\Omega_{K*}$, where 
$\Omega_{K*}=\sqrt{GM/R_*^3}$.
   For most of the figures
$\omega_*=0.5$, and this corresponds to
a period of $P_*=2\pi/\Omega_*
\approx 685$ s.  
   
    Using the mass accretion rate
of equation (11) and $\mu_{30}=1$,
the magnetospheric
radius of the non-rotating star, from equation (1), is
$r_{m0} \approx 6.6\times 10^{10}~{\rm cm}
\approx 8 R_*$.
   The magnetospheric radius of the rotating
star with $\omega_*=0.5$, from equation (2),  is
$r_{m\Omega} \approx 3.1\times 10^{10}~{\rm cm}
\approx 3.8R_*$.
  From Figure 8 note  that the equatorial profile
of the angular rotation rate $\Omega(r,0)$
for $\omega_*=0.5$
is approximately flat out to 
$r_{flat}\approx 4.2R_*$ which is roughly the
same as $r_{m\Omega}$.
   As required,
the corotation radius, $r_{cor}=R_*/\omega_*^{2/3}
\approx 1.6 R_*$, is appreciably less than $r_{m\Omega}$.
    For  smaller values of $\omega_*$ in
Figure 8,  $r_{flat}$ decreases gradually.  
  This dependence on $\omega_*$  differs from that
of equation (2) which may be due to the fact
that $r_{m\Omega}$ is not very large compared
with $r_{cor}$.
    As $\omega_*$ decreases,
$r_{cor}$ increases rapidly.  
Thus the case $\omega_*=0.1$, where 
$r_{cor} >r_{flat}$, is not in the
propeller regime.

  For  the considered case of
$\beta=10^{-7}$, $b_0=10$, 
$\tilde\eta_m=10^{-5}$, and 
$\omega_*=0.5$, we have evaluated
equation (9) numerically with
the result that $\dot{L} \approx
-4.7\times 10^{28}~{\rm g (cm/s)}^2$.
    The theoretical estimate
of $\dot{L}$ from equation (3)
for the mentioned values of $\mu$
and $r_{m\Omega}$ is
$\dot{L}_{th} = 1.5\times 10^{28}~
{\rm g (cm/s)}^2$.
      A small, $70\%$ reduction of $r_{m\Omega}$
brings $\dot{L}_{th}$ into coincidence
with the $\dot{L}$ from the simulations.

\section{Conclusions}

  Axisymmetric magnetohydrodynamic 
simulations of
Bondi accretion to a rotating
magnetized
star in the propeller
regime of accretion  have
shown that: 
     (1) A new regime of matter
flow forms around
a rotating star.
   Matter falls down along
the axis, but only a small
fraction of the incoming
matter accretes to the
surface of the star. 
   Most of the matter is expelled 
radially in the
equatorial plane by the
rotating magnetosphere of the
star. 
   A low-density 
torus forms in the
equatorial region which
rotates with velocity significantly
larger than the radial velocity.
   Large scale vortices form above
and below the equatorial plane.
  (2) The star is spun-down by
the  magnetic torque and to a lesser
extent the matter
torque at its surface.
   The rate of loss of angular momentum $\dot{L}$
is proportional to $-\Omega_*^{1.3}\mu^{0.8}$,
and it is approximately independent of $\eta_m$.
This dependence differs from the 
predicted dependence of equation (3) probably because
the corotation radius is not much smaller
than the magnetospheric radius $r_m$.
  The rotational energy lost by 
the star goes into the directed
and thermal energy of plasma.
    (3) The accretion rate
to the star
is much less than the Bondi 
accretion rate and 
decreases as (a) the
star's rotation rate increases 
($\propto \Omega_*^{-1.0}$), 
(b) as the star's magnetic moment
increases
($ \propto \mu^{-1.7}$), and as the 
magnetic diffusivity
decreases [$\propto (\eta_m)^{0.4}$].
  (4) Because the
accretion rate to the star is
less than the Bondi rate, a shock wave
forms in our simulations
 and propagates outward.
   It has the
shape of an ellipsoid flattened along
the rotation axis of the star.

\acknowledgments
This work was supported in part by NASA grant
NAG5-9047, by NSF grant AST-9986936,
and by Russian program ``Astronomy.''
 M.M.R. is grateful to an NSF POWRE 
grant for partial support.
R.V.E.L. was partially supported 
by grant NAG5-9735.
We thank Dr. V.V. Savel'ev 
for the development of the 
original version of the MHD
code used in this work.


\begin{references}

\reference{aro76} Arons, J.,
 \& Lea, S.M. 1976,
ApJ, 207, 914



\reference{bon52} Bondi, H. 1952, MNRAS, 112, 195

\reference{} Cui, W. 1997, ApJ, 482, L163

\reference{dav73} Davidson, K.,
\& Ostriker, J.P. 1973, ApJ, 179,
585

\reference{} Davies, R.E., Fabian, A.C.,
\& Pringle, J.E. 1979,
MNRAS, 186, 779

\reference{} Davies, R.E., \& Pringle, J.E. 1981, 
MNRAS, 196, 209


\reference{} Goldreich, P., \& Julian, W.H. 1969, 
ApJ, 157, 869



\reference{ill75} Illarionov, A.F.,
 \& Sunyaev, R.A. 1975, A\&A, 39, 185



\reference{lip92} Lipunov, V.M. 1992,
{\it Astrophysics of Neutron Stars},
(Berlin: Springer Verlag)

\reference{lov95} Lovelace, R.V.E., Romanova, M.M., \&
Bisnovatyi--Kogan, G.S. 1995, MNRAS, 275, 244

\reference{} -------------
 1999, ApJ, 514, 368

\reference{} -------------
 2002, in preparation

\reference{} Nelson, R.W., Salpeter, E.E., \&
Wasserman, I. 1993, ApJ, 418, 874






\reference{} Romanova, M.M., Ustyugova, G.V., Koldoba, A.V., \&
Lovelace, R.V.E. 2002, ApJ, in press

\reference{ruf94a} Ruffert, M. 1994a,
\apj, 427, 342

\reference{ruf94b} Ruffert, M. 1994b,
Astron. Astrophys. Suppl. Ser. 1994,
106, 505



\reference{} Shvartsman, V.F. 1970, Radiofizika, 13, 1852

\reference{sha83} Shapiro, S.L., \& Teukolsky, S.A. 1983, ``Black
holes, white dwarfs, and neutron stars",  (Wiley-Interscience)


\reference{} Stella, L., White, N.E.,
\& Rosner, R. 1986, ApJ,
308, 669



\reference{tor99} Toropin, Yu.M., 
Toropina, O.D., Savelyev, V.V.,
Romanova, M.M., Chechetkin, V.M., 
\& Lovelace, R.V.E. 1999,
ApJ, 517, 906

\reference{tor01} Toropina, O.D.,
Romanova, M.M., Toropin, Yu.M., 
\& Lovelace, R.V.E. 2001,
ApJ, 561, 964

\reference{tor01} Toropina, O.D.,
Romanova, M.M., Toropin, Yu.M., 
\& Lovelace, R.V.E. 2002,
in preparation

\reference{tre91} Treves, A., 
\& Colpi, M. 1991, A\&A, 241, 107

\reference{tre00} Treves, A. 2000, Turolla, R., Zane, S., 
\& Colpi, M. 2002, PASP, 112, 769

\reference{} Wang, Y.-M., 
\& Robertson, J.A. 1985, A\&A 151, 361

\reference {zhu93} Zhukov, V.T.,
Zabrodin, A.V., \& Feodoritova,~O.B. 1993,
Comp. Maths. Math. Phys.,
33, No. 8, 1099


\end{references}
\end{document}